\documentclass{ifacconf}

\usepackage{graphicx}      
\usepackage{natbib}        
\usepackage{amsmath}
\usepackage{siunitx}
\usepackage{amsfonts}

\begin{document}
\begin{frontmatter}

\title{Effect of State of Charge Uncertainty on Battery Energy Storage Systems}

\author[First]{Sonia Martin}  
\author[Second]{Simona Onori} 
\author[Third]{Ram Rajagopal}

\address[First]{Department of Mechanical Engineering, Stanford University, 
   Stanford, CA 94305 USA, email: soniamartin@stanford.edu }
\address[Second]{Department of Energy Resources Engineering, Stanford University, 
   Stanford, CA 94305 USA, email: sonori@stanford.edu}
\address[Third]{Department of Civil and Environmental Engineering, Stanford University, Stanford, CA 94305 USA, email: ramr@stanford.edu}

\begin{abstract}                
Battery energy storage systems (BESSs) provide many benefits to the electricity grid, including stability, backup power, and flexibility in introducing more clean energy sources. As BESS penetration grows, knowledge of the uncertainty in the battery's state of charge (SOC) estimate is crucial for planning optimal BESS power injection trajectories. This paper proposes a framework for quantifying SOC estimation uncertainty based on battery rest periods. An uncertainty analysis is presented for a BESS participating in the frequency regulation market.
\end{abstract}


\end{frontmatter}

\section{Introduction}

Electricity is essential for communities to thrive but its generation and distribution are often expensive and are major contributors to global warming. While high penetration of intermittent clean energy sources significantly reduces electricity grid emissions, these sources also introduce grid instabilities. Battery energy storage systems (BESSs), defined as grid-connected, stationary battery packs, not only stabilize the grid but also ensure continuous power for consumers. Thus, expanding BESS installation will decrease greenhouse gas emissions by enabling a higher percentage of clean energy in the grid. For a BESS to adequately profit from the wholesale electricity market, however, it must be able to accurately bid hourly power and capacity levels, which is only possible with an accurate state of charge (SOC) estimate. In this work we develop a novel SOC uncertainty quantification methodology based on rest periods and voltage relaxation dynamics. We begin by motivating the work, then we describe the SOC uncertainty model, and lastly we present results and analysis of SOC uncertainty from two experimental current profiles.

\section{Motivation and Background}

While the literature on SOC estimation algorithms is expansive, many works do not provide uncertainty bounds on their SOC estimates. Errors in SOC estimation can arise from the model used to represent the electrochemical battery cell as well as from physical sensor imprecision and inaccuracy. Research by \cite{investigating_errors} surveys various errors in SOC estimation, including errors present in equivalent circuit models (ECM), a class of battery models used in this work. One prominent estimation error source is battery voltage relaxation dynamics: when disconnected from a load, the battery cell takes minutes to hours to reach an equilibrium voltage state, or the open circuit voltage (OCV). For most commercially available lithium-ion batteries, SOC can be accurately estimated from OCV once the battery reaches equilibrium; however, error is introduced during the relaxation phase. Furthermore, sensor errors in current and voltage measurements have a significant effect on SOC estimation error, especially in systems with low-resolution sensors \citep{dubarry}. Specifically, current sensor error can grow in the presence of high temperatures (shunt resistor sensor) or have an offset at zero current (Hall-effect sensor) \citep{wei_future_2021}.

Works such as \cite{dubarry} and \cite{zhu_soc_2021} provide experimental SOC error analysis, and only a few works offer a mathematical justification or uncertainty prediction. Notably, \cite{lin_theoretical_2018} discusses error resulting from sensor bias and variance with the Kalman filter. \cite{chalmers} derives an uncertainty prediction framework based on voltage and current sensor error in the context of hybrid electric vehicle energy management with SOC feedback. While both of these works take into account sensor errors, they do not incorporate temporal errors associated with voltage relaxation dynamics. Similar to the voltage-based SOC update step proposed in \cite{chalmers}, \cite{tesla_patent} formulates a SOC estimation method with an update step and includes temporal errors as well as temperature-dependent dynamics. \cite{mitigating} also includes dynamic error proportional to the rate of voltage relaxation. Our paper builds upon these works to provide a SOC uncertainty methodology that addresses sensor errors as well as temporal errors, which, to the best of our knowledge, has not been seen in the literature.

Quantifying SOC uncertainty is useful for many applications, including electric vehicle (EV) charge management and BESS power injection forecasting. Works such as \cite{ghosh_effect_2020} and \cite{padhee_fixed-flexible_2020} utilize knowledge of SOC estimation uncertainty to optimally control battery behavior in hybrid electric vehicles and stationary BESSs, respectively. For the stationary storage case, especially in networked or virtualized systems when multiple physical batteries may be aggregated, understanding potential uncertainties ensures the controllers do not implement overly conservative planning algorithms. Overall, studies on BESSs are beginning to incorporate system uncertainties, but significant research is still needed to integrate detailed battery models with SOC uncertainty.

\section{SOC Uncertainty Model}
This section describes the battery model, parameter identification procedure, model errors, and SOC estimation algorithm used in the uncertainty quantification framework. Variable definitions are listed in Table~\ref{tab:vars}.

\begin{table}[h]

 \begin{center}

   \caption{ Variable Descriptions
}\label{tab:vars}
   \begin{tabular}{c c }
     \hline
     Variable & Description \\
     \hline
     $SOC$ & True state of charge \\
     $\eta$ & Coulombic efficiency  \\
     $\Delta t$ &   Timestep size [s]\\
     $Q$ & Capacity [Ah]  \\
     $i$ &   Current [A]\\
     $R_0$ & Series resistance [$\Omega$]  \\
     $R_1$ &  RC pair parallel resistance [$\Omega$]\\
     $C_1$ &  RC pair parallel capacitance [F]\\
      $V_{C1}$ & Voltage across $C_1$ [V]  \\
     $V$ &  Terminal voltage [V]\\
$OCV$ &  Open circuit voltage [V]\\
$\tilde{SOC}$ &  Open loop estimated SOC via voltage inversion \\
 $\hat{SOC}$  &  Closed loop estimated SOC  \\
$\mu, \alpha, \beta, \lambda_1, \lambda_2$ &  Error parameters \\
$\delta$ &  Feedback gain\\
$e$ &  SOC error \\
$u$ & Uncertainty (error standard deviation) \\
     \hline 
   \end{tabular}

 \end{center}
\end{table} 

\subsection{Equivalent Circuit Model}

\begin{figure}[t]
\vspace{-2.5em}
\includegraphics[scale=.9]{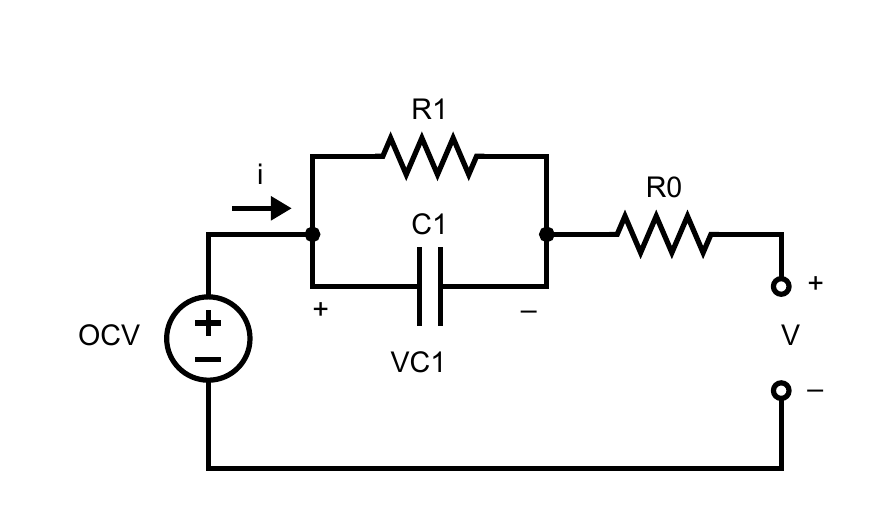}
\vspace{-.5em}
\caption{Equivalent circuit model depicting (from left to right) OCV, RC pair modelling transient voltage, series resistance, and terminal voltage. }
\label{ECM}
\end{figure}

The true SOC is simulated via Coulomb counting and the terminal voltage is simulated with an equivalent circuit model (ECM), shown in  Fig.~\ref{ECM} \citep{plett_extended_2004}. The ECM is chosen because it is flexible and feasible to run on a commercial BMS. Furthermore, BESSs often operate with low C-rates, so this paper utilizes a one resistor-capacitor pair ECM for a proof of concept for SOC uncertainty. For applications that require higher C-rates, a higher order ECM is advisable~\citep{ahmed_model-based_2015}. In this work we assume a BESS can be approximated as a single lumped cell with measured pack level voltage and current, as is done in \cite{rosewater_battery_2019}. The three coupled and discretized ECM equations are as follows.
\begin{align}
    SOC_{k+1}=SOC_k-\frac{\eta\Delta t_k}{Q} i_k \label{CC} \\
    V_{C_1,k+1}=V_{C_1,k} \exp \left(\frac{-\Delta t}{R_1C_1} \right) \notag \mspace{50mu} \\ 
      +R_1\left( 1-\exp\left(\frac{-\Delta t}{R_1C_1}  \right) \right)i_k   \label{VC1} \\
    V_k=OCV\left(SOC_k\right)-V_{C_1,k}-R_0i_k \label{KVL}
\end{align}
Eq.~\eqref{CC} is the true SOC via Coulomb counting, where $\Delta t_k$ is the sampling time, i.e. $\Delta t_k = t_{k+1}-t_k$. Discharge current is positive. Eq.~\eqref{VC1} updates the voltage across the capacitor at each timestep as an exponentially decaying function of the prior voltage and the current. Finally, Eq.~\eqref{KVL} is from Kirchoff's Voltage Law, where OCV is a nonlinear function of SOC, i.e. $OCV=f\left( SOC\right)$, as shown in the blue curve in Fig.~\ref{dSOCdOCV}.  

\subsection{Parameter Identification}

The parameters $R_1$, $C_1$, and $R_0$ are functions of SOC and are identified via the gradient-free genetic algorithm for function minimization. The hybrid pulse power characterization (HPPC) test data from a Nickel Manganese Cobalt (NMC) Li-ion cell from \cite{batt_data} is used in this work. While our work does not explicitly account for the temperature-dependency of the parameters, future work will include this dependency especially as it relates to the heterogeneity of cell temperatures within a battery pack.

\subsection{Model Errors}

Before deriving the SOC uncertainty algorithm, we first define each modeled error source and corresponding error distribution. Both the current and voltage sensor errors are modeled as normal distributions, as is done in \cite{zhao_observability_2017}. While the current sensor error includes a bias in our model, the voltage sensor error is assumed to be zero-mean. If the sensor is specifically known to have a bias, it can be calibrated out before the model is used.

The normally distributed current sensor error has mean equal to the sensor drift, and its standard deviation is representative of both random noise and uncertainty in the drift as shown below: 
\begin{align}
     \epsilon_{i,k} \sim \mathcal{N}\left(\mu,\alpha+\beta  i_k^2\right) 
\end{align}

$\mu$ is the sensor drift, $\alpha$ is a small constant introduced to avoid singularities, and $\beta$ is a coefficient to determine the variance of the bias plus the noise as a function of the current magnitude. These parameters are fixed in simulation, and $\mu$ and $\beta$ are found from the testing device specifications (see Table~\ref{tab:params}) \citep{sensor_drift, arbin}.

There are two voltage errors introduced in the voltage-based SOC calculation, shown below, which is an inversion of the tracked OCV. 
\begin{align}
    \Tilde{SOC}_k=f^{-1}\left( V_k +V_{C_{1},k}\right)=SOC_k+\epsilon_{V,1,k}+\epsilon_{V,2,k} \label{soc_tilde}
    \end{align}

\begin{figure}[t]
\includegraphics[scale=.53]{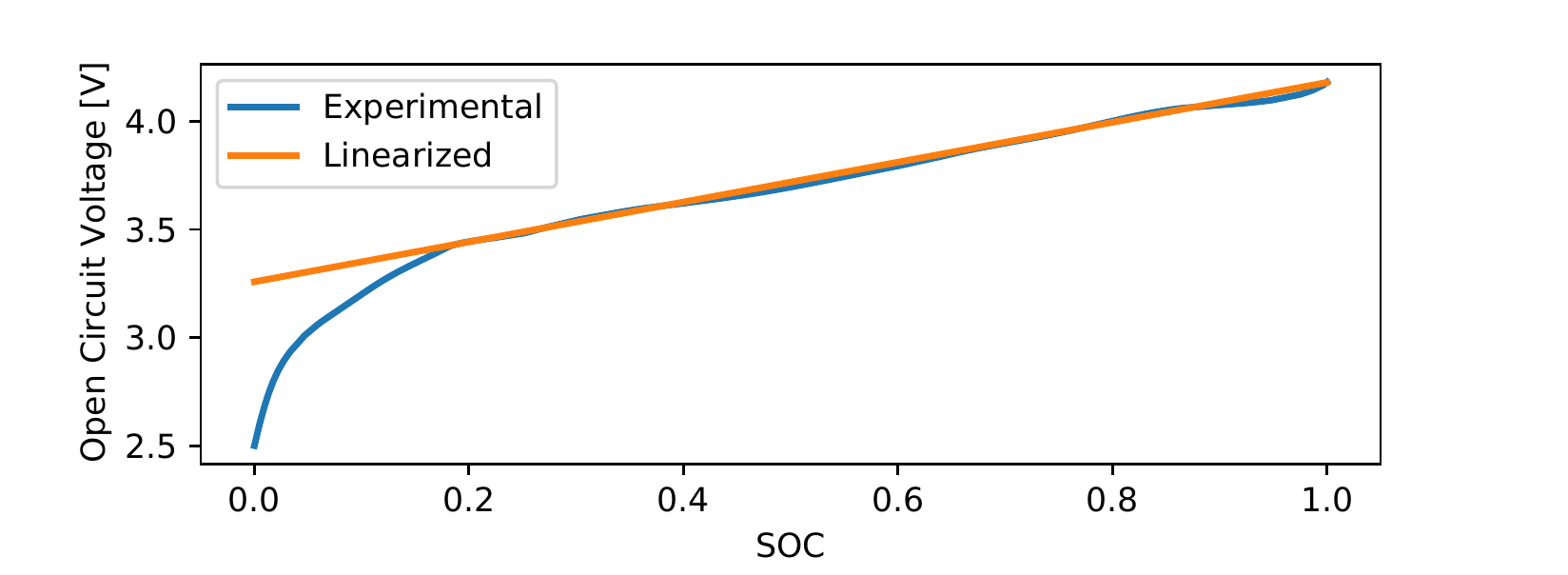}
\caption{OCV-SOC curve for voltage inversion to SOC; the experimental curve (nonlinear) and the linearized curve are shown. }
\label{dSOCdOCV}
\end{figure}

This is derived from rearranging Eq.~\eqref{KVL} and assuming $i_k=0$ for this inversion, since this measurement is used only during rest periods. The dynamics of $\epsilon_{V,1,k}$ are characterized by the following statistical distribution: 
  \begin{align}
        \epsilon_{V,1,k} \sim\mathcal{N}\left(0, \lambda_1 \frac{d\hat{SOC}_k}{dOCV} \right) \label{ev1}  
    \end{align}
In Eq.~\eqref{soc_tilde}, a small error in the calculated $OCV_k=V_k +V_{C_{1},k}$ input will produce an error, $\epsilon_{V,1,k}$, in the SOC output. By linearizing the OCV-SOC curve, the slope $\frac{dSOC}{dOCV}$ becomes constant. Fig.~\ref{dSOCdOCV} shows the linearization. This work examines only SOC values above 20\% where the linearization is a good approximation of the true curve, as many BMSs restrict the SOC of operation to values greater than 20\% to reduce aging. The model could extend to SOC values below 20\% by using a piecewise linear approximation of the curve.

The second voltage error results from the inability to determine the SOC from the OCV until the terminal voltage has reached steady state. The dynamic distribution of $\epsilon_{V,2,k}$ is a function of the time constant, $R_1C_1$ and the time at rest:
\begin{align}
        \epsilon_{V,2,k}\sim \mathcal{N}\left(0,\lambda_2 \frac{R_1 C_1}{t_R}\right)  \label{ev2}
\end{align}
$t_R$, the elapsed time at rest (a period of time when $i_k=0$), is defined as $t_R=t_k-t_{\textrm{rest}}$. $t_{\textrm{rest}}$ is the time stamp when $i_k=0$ that begins the rest period. Eq.~\eqref{ev2} avoids singularities because this error is not introduced unless $t_R>0$.

\begin{table}[h]

 \begin{center}

   \caption{ Model Parameter Values
}\label{tab:params}
   \begin{tabular}{c c }
     \hline
     Parameter & Value  \\
     \hline
     $Q$ & \SI{4.85}{Ah}   \\
     $\eta$ & 1.0  \\
     $\mu$ & \SI{.03}{A}  \\
     $\alpha$ &  \SI{1e-7}{A} \\
      $\beta$  &  \SI{1.4e-4}{\frac{A}{A^2}}  \\
      $\lambda_1$ & \SI{1e-6}{V^2}  \\
      $\lambda_2$ & \SI{6e-7}{V} \\
     \hline 
   \end{tabular}

 \end{center}
\end{table} 

\subsection{Uncertainty Quantification}
The SOC uncertainty is derived from a SOC estimation algorithm that combines Coulomb counting (integrating net current flowing into and out of the battery) with a correction term based on the error between the prior SOC estimate and the inversion of the terminal voltage measurement ($\tilde{SOC}$). The algorithm is adapted from work by \cite{chalmers}.

In dynamic mode ($i \neq 0$), Coulomb counting is used for SOC estimation, and at rest ($i=0$) the SOC is updated based on the voltage measurement. The gain for the update step when the battery is at rest is a function of the uncertainty at each time step and is derived in the following equations.

\begin{align}
    \hat{SOC}_{k+1}=\left(1-\delta_k\right)\hat{SOC}_{k}-\frac{\eta\Delta t_k}{Q} ( i_k+\epsilon_{i,k})+\delta_k  \Tilde{SOC_k} \\
    \delta_k=      \begin{cases} 
      0 & i_k \neq 0 \\
     \frac{u_k^2}{u_k^2+\lambda_1 \frac{d\hat{SOC}_k}{dOCV}  +  \lambda_2 \frac{R_1 C_1}{t_R}}  & i_k=0 \\
   \end{cases} \notag 
   \end{align}

\begin{figure}[ht]
\includegraphics[scale=.26]{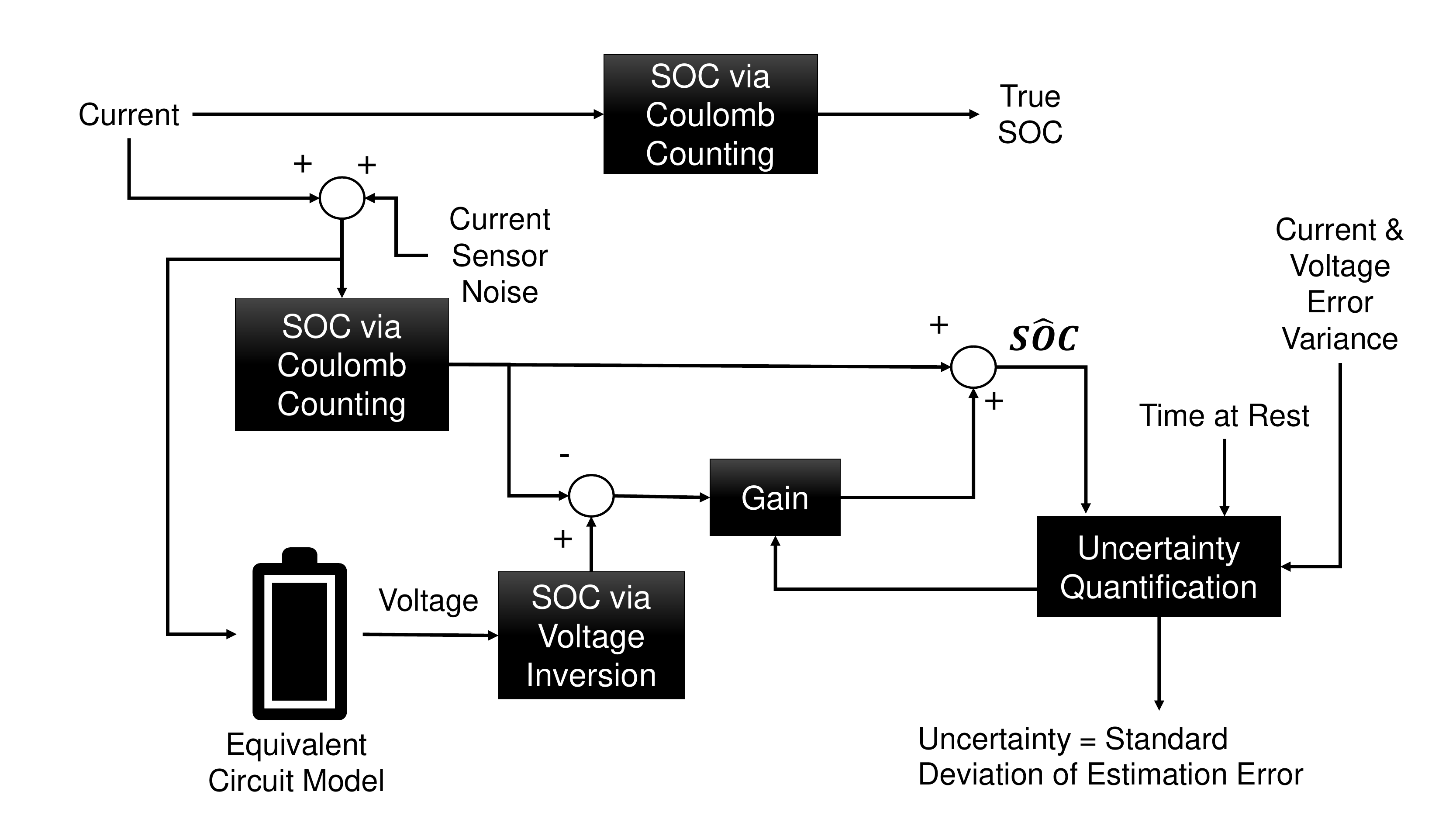}
\caption{The block diagram for the closed loop SOC estimation and uncertainty quantification algorithm. The true SOC from error-free Coulomb counting is unknown in real BMS operation. $\hat{SOC}$ is updated by inversion of the OCV; the update step gain is tuned from the uncertainty.}
\label{diagram}
\end{figure}

 Fig.~\ref{diagram} displays the block diagram for the uncertainty quantification and SOC estimation process. At rest, SOC estimation is a closed-loop feedback control system where the gain is a function of the uncertainty. Specifically, the estimation uncertainty is defined as the standard deviation of the estimation error. Although the true error is unknown, uncertainty can be calculated from the known distributions of the error sources.

The estimation uncertainty derivation is as follows, where $e$ is the error and $u$ is the uncertainty.
   \begin{align}
   e_{k+1} &= SOC_{k+1}-\hat{SOC}_{k+1} \label{e1}\\
   & =SOC_k-\frac{\eta\Delta t_k}{Q} i_k 
   - \hat{SOC}_{k}  \notag\\
  &  +\frac{\eta\Delta t_k}{Q} ( i_k+\epsilon_{i,k})-\delta_k  ( \Tilde{SOC_k}-\hat{SOC}_k ) \label{e2}\\
 & =   SOC_k-\hat{SOC}_{k}+\frac{\eta\Delta t_k}{Q}\epsilon_{i,k}    -\delta_k SOC_k  \notag\\ 
 & +\delta_k\hat{SOC}_{k}
 -\delta_k(\epsilon_{V,1,k}+\epsilon_{V,2,k}) \label{e3}  
   \end{align}

   Eq.~\eqref{e2} expands $\hat{SOC}_{k+1}$ in the error equation and Eq.~\eqref{e3} simplifies the expression and replaces $\tilde{SOC}_{k}$ by its definition in Eq.~\eqref{soc_tilde}. We then replace $SOC_{k}-\hat{SOC}_{k}$ with $e_{k}$,  combine like terms, and flip the voltage error sign for mathematical simplicity given that it is normally distributed.
   
   \begin{align}
  e_{k+1} = (1-\delta_k)e_k +\frac{\eta\Delta t_k}{Q}\epsilon_{i,k}+\delta_k(\epsilon_{V,1,k}+\epsilon_{V,2,k}) \label{e4}   \\
   u_{k+1}^2=\textrm{Var}[e_{k+1} ] \label{e5}
  \end{align}
  The square of the uncertainty is equal to the variance of the error as shown in Eq.~\eqref{e5}. To derive this, we use the known variance of each added error source: current sensor error, voltage inversion error, and voltage relaxation error. $\delta_k$, the feedback gain, is selected to be $0$ when $i\neq0$ since voltage inversion to find SOC is not meaningful during dynamic mode, but when $i=0$, $\delta_k$ is set as the gain that minimizes the uncertainty at that timestep.
  
  \begin{align}
  u_{k+1}^2=(1-\delta_k)^2\textrm{Var}[e_k]+\frac{\eta^2{\Delta t_k}^2}{Q^2}\textrm{Var}[\epsilon_{i,k}]\notag \mspace{50mu}\\
  +\delta_k^2\textrm{Var}[\epsilon_{V,1,k}] +\delta_k^2\textrm{Var}[\epsilon_{V,2,k}] \label{e6} 
    \end{align}

Eq.~\eqref{e6} defines the uncertainty stochastically. In simulation, we introduce the assumption that the variances of the errors are known. This allows us to calculate the uncertainty deterministically. That is to say, we replace the variance with its expected value:  $\textrm{Var}[\epsilon_{i,k}]=\mathbb{E}[\epsilon_{i,k}^2]=\alpha+\beta i_k^2 $ and so on for each error source.

   \begin{align}
     u_{k+1}^2=(1-\delta_k)^2u_k^2+\frac{\eta^2{\Delta t_{k}}^2}{Q^2}(\alpha+\beta i_k^2)\notag \mspace{50mu}\\
    +\delta_k^2\left( \lambda_1 \frac{d\hat{SOC}_k}{dOCV}  +    \lambda_2 \frac{R_1 C_1}{t_R}\right)\label{e7} 
          \end{align}

    \begin{align}
    =u_k^2-2\delta_k u_k^2+\delta_k^2 \left(u_k^2+\lambda_1 \frac{d\hat{SOC}_k}{dOCV}  +  \lambda_2 \frac{R_1 C_1}{t_R}\right) \notag \\
    +\frac{\eta^2{\Delta t_{k}}^2}{Q^2}(\alpha+\beta i_k^2)\label{e8}
    \end{align}
    From Eq.~\eqref{e8}, a simplified uncertainty expression is derived. When $i\neq0$, the square of the uncertainty at each time step is equal to its prior value plus the variance of the current error, as follows:
    
    \begin{align}
     u_{k+1}^2=u_k^2
    +\frac{\eta^2{\Delta t_{k}}^2}{Q^2}(\alpha+\beta  i_k^2),~i_k\neq0\label{e9}
    \end{align}
    When $i_k=0$, $\delta_k$ can be selected to minimize the uncertainty, given that $u_k^2$ and the voltage error variance are greater than zero, as follows: 
    \begin{align}
    \delta_{k|i_k=0}=\notag \mspace{350mu}\\ 
    \underset{\delta_k}{\mathrm{argmin}}\ \delta_k^2 \left(u_k^2+\lambda_1 \frac{d\hat{SOC}_k}{dOCV}  +  \lambda_2 \frac{R_1 C_1}{t_R}\right) -2\delta_k u_k^2 \notag\\
    =\frac{u_k^2}{u_k^2+\lambda_1 \frac{d\hat{SOC}_k}{dOCV}  +  \lambda_2 \frac{R_1 C_1}{t_R}} \label{e10}
\end{align}

Since the current and voltage errors are normally distributed, a 95\% confidence interval can be constructed around the estimated SOC plus the cumulative bias. In dynamic operation, the bias evolves according to:
\begin{align}
\textrm{bias}_{k+1}=\textrm{bias}_k+\frac{\eta{\Delta t_{k}}}{Q}\mu ,~i_k\neq0 \label{bias}
\end{align}
At rest, the bias  is reset and decreases linearly proportional to the time-based voltage error as shown below.
\begin{align}
\textrm{bias}_{k+1}=\textrm{bias}_k\lambda_2 \frac{R_1C_1}{t_R} ,~i_k=0 \label{bias2}
\end{align}
The uncertainty bounds are formulated as:
\begin{align}
\textrm{95\% Confidence Interval}=\hat{SOC}_k+\mu_k \pm 2u_k \label{e11}
\end{align}

\section{Results}

This section includes two simulation experiments. The first is a simple, illustrative example of the decrease in uncertainty from the update step when the battery is at rest. The second provides an uncertainty analysis for a battery participating in frequency regulation. Each experiment is simulated on the aforementioned 4.85Ah NMC cell.

\subsection{Uncertainty for ``At Rest'' Experiment}

\begin{figure}[htbp]
$\mspace{20mu}$
\includegraphics[trim={0 0 4em 0},clip, scale=.53]{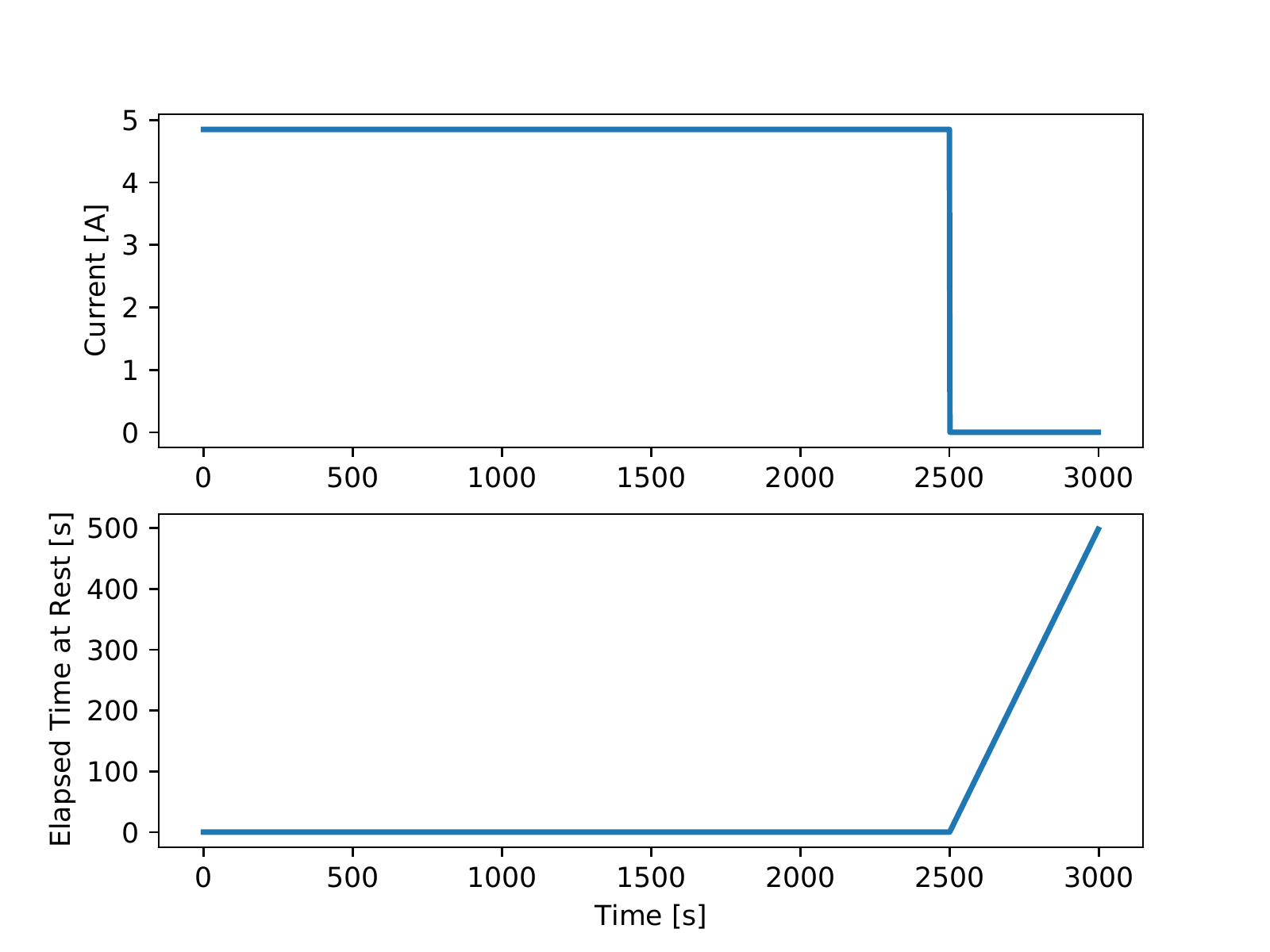}
\caption{(Top) current profile and (bottom) the associated elapsed time at rest, $t_R$, for a 1C constant discharge. The cell discharges at 1C (4.85A) for 2500 seconds; once at rest, current is zero for the remaining 500 seconds. The elapsed time at rest tracks the length in seconds of each uninterrupted rest period, which in this profile is the last 500 seconds. }
\label{rest}
\end{figure}

When the battery cell is at rest after charging or discharging, the estimated SOC is updated using the voltage measurement; this experiment demonstrates how the uncertainty decreases during a rest period after a constant discharge. We examine a current profile from a 100\% initial SOC consisting of a 1C discharge for 2500 seconds followed by 500 seconds rest, with a 1 second sampling time, as shown in Fig.~\ref{rest}. 

\begin{figure}[htbp]
$\mspace{20mu}$
\includegraphics[trim={0 0 4em 0},clip, scale=.53]{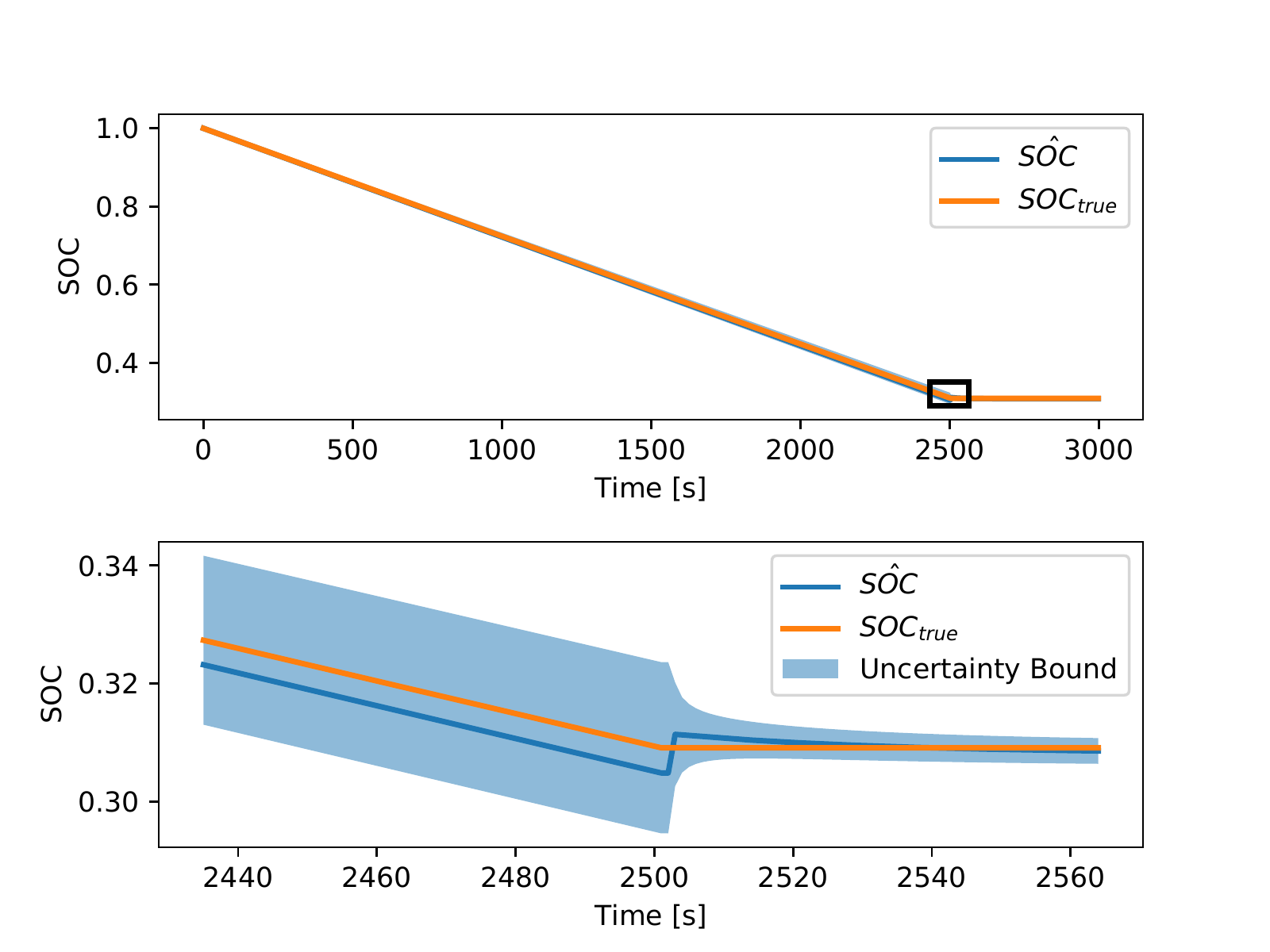}
\caption{The SOC decreases linearly until it stays constant after $t=2500$ (top). A magnified view of the SOC (bottom) shows the the jump in $\hat{SOC}$ at rest and the corresponding decrease in uncertainty bounds.}
\label{soc_const}
\end{figure}
\begin{figure}[htbp]
$\mspace{20mu}$
\includegraphics[trim={0 0 5em 0},clip, scale=.53]{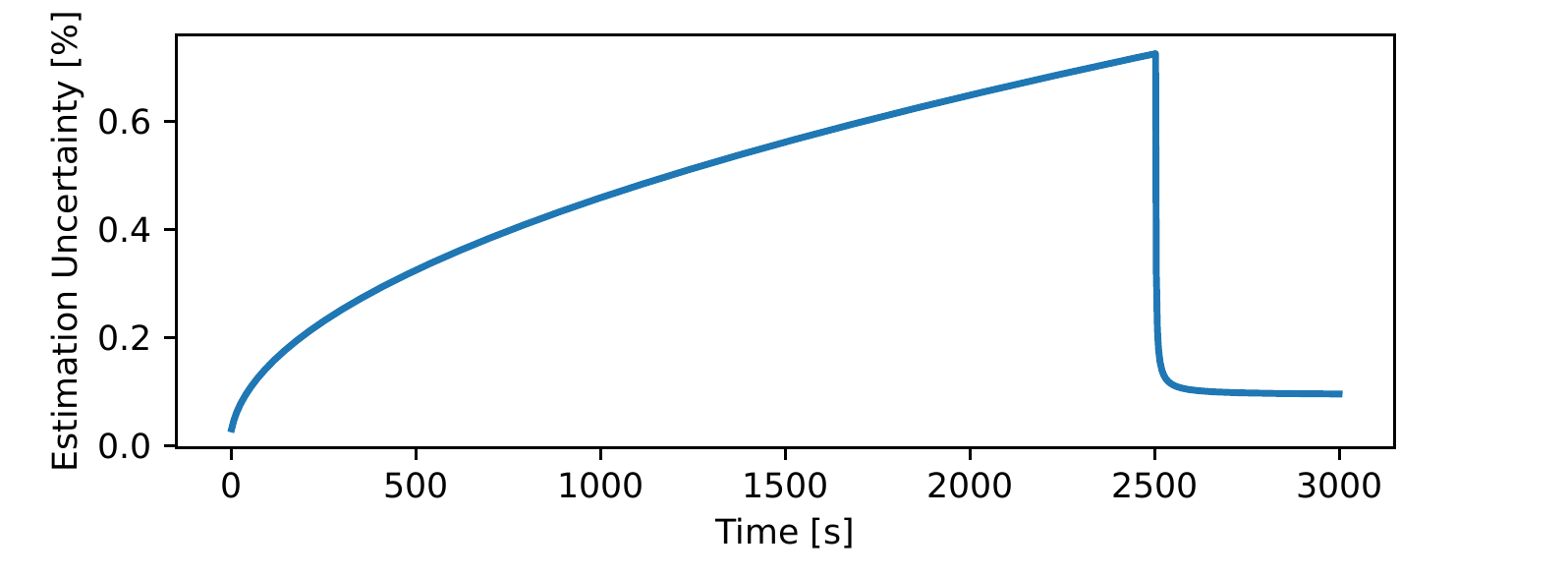}
\caption{The estimation uncertainty increases as the current error grows during dynamic operation. At rest, the voltage update significantly decreases the uncertainty as the voltage relaxes. }
\label{u_const}
\end{figure}

Fig.~\ref{soc_const} displays the SOC update at rest and shows how the estimated SOC begins to converge toward the true SOC. Lastly, the uncertainty $u$ is shown in Fig.~\ref{u_const}.

\subsection{Uncertainty for Frequency Regulation Experiment}
\begin{figure}[htbp]
$\mspace{10mu}$
\includegraphics[trim={0 0 2em 0},clip, scale=.53]{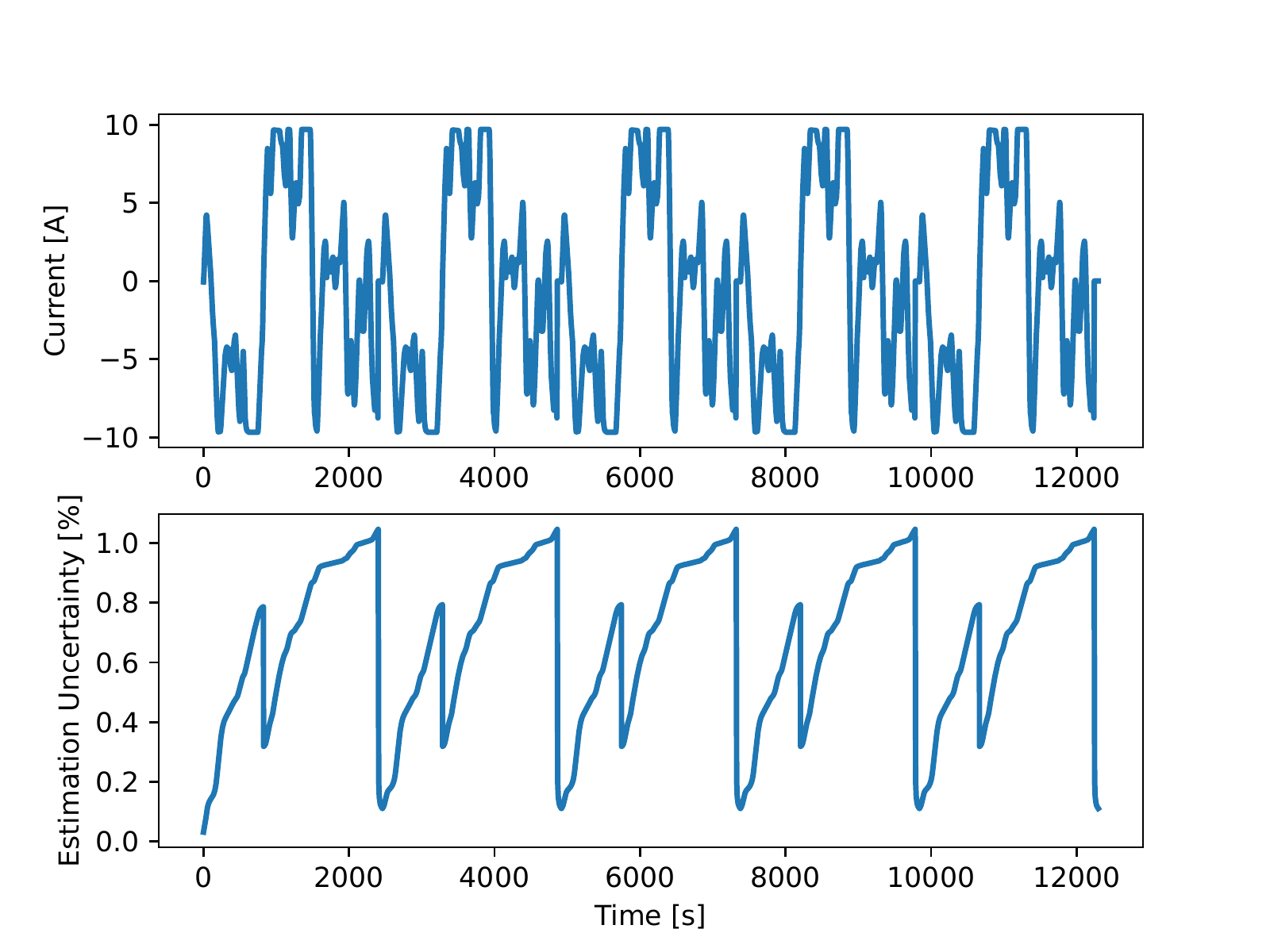}
\caption{Current profile (top) and estimation uncertainty (bottom) for five concatenated RegD signals. After the first signal, the uncertainty is periodic: the brief rest within the signal ensures the uncertainty doesn't grow unbounded, but the lack of a longer rest means that the uncertainty never drops below 0.25\%.}
\label{freq1}
\end{figure}

\begin{figure}[htbp]
$\mspace{10mu}$
\includegraphics[trim={0 0 5em 0},clip, scale=.52]{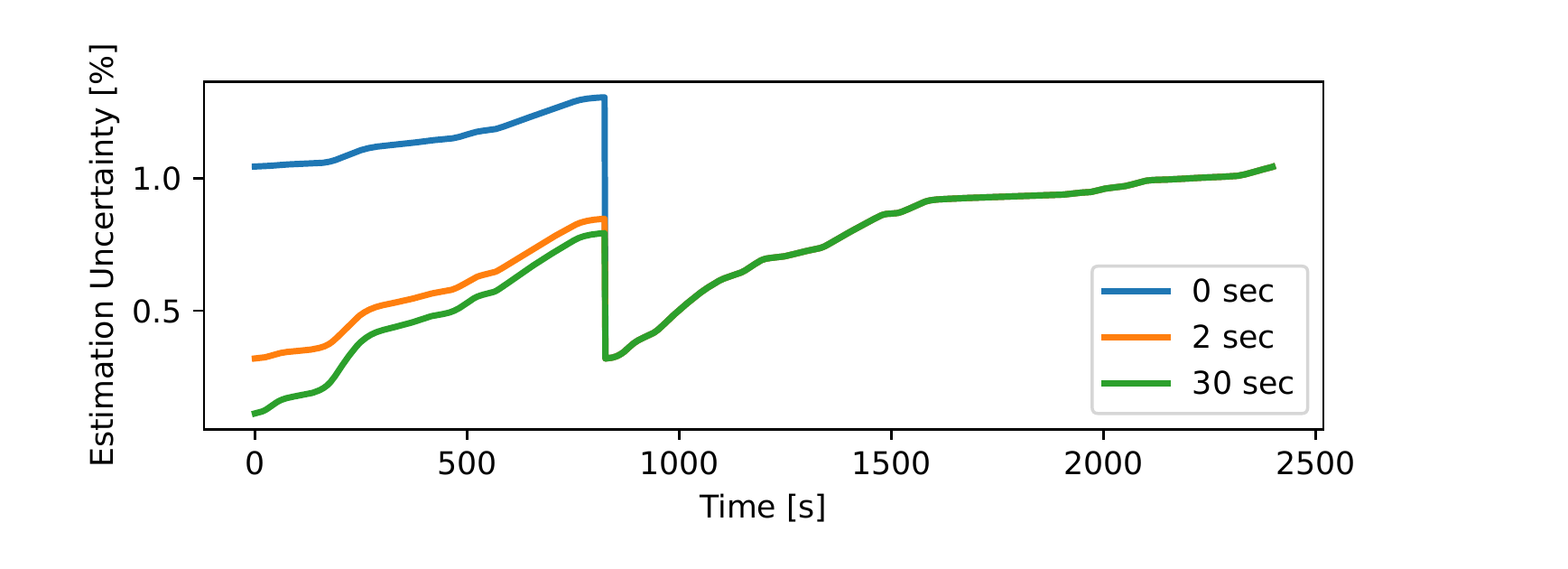}
\caption{This plot shows the uncertainty during the last regulation signal out of a sequence of five signals. Each curve illustrates the uncertainty for a different length of rest between each RegD signal, starting with no rest and increasing to 30 seconds of rest in between. Even a 2 second rest will decrease the uncertainty at the start of the signal, but a rest of at least 30 seconds ensures that the uncertainty decreases to below 0.1\% at the start of each signal. }
\label{freq2}
\end{figure}

The uncertainty analysis for frequency regulation participation has two aims. First, we quantify the uncertainty arising from repeated, concatenated frequency regulation signals. Second, we determine how much rest time in between signals is necessary to continue following the signal while remaining under 0.5\% uncertainty. The specific frequency regulation signal analyzed is the Regulation D (RegD) signal, a dynamic signal used to maintain grid frequency. The RegD signal changes value every 2 seconds, lasts for 40 minutes, and is scaled such that the maximum current magnitude does not exceed 2C. The signal data is a sample from the PJM ISO \citep{PJM}.  Fig.~\ref{freq1} shows the current profile for five concatenated RegD signals and the corresponding estimation uncertainty. The brief, 4 second zero crossing during the signal allows the estimation uncertainty to drop during operation, but without a designated rest and calibration period between each 40 minute signal, the uncertainty never drops below 0.25\%. The effect of adding a rest period between signals is shown in Fig.~\ref{freq2}.

\section{Discussion}
Analyzing the SOC uncertainty illuminates a fundamental aspect of SOC estimation: without a rest period, the current drift will eventually lead to a high enough uncertainty to restrict power trajectory planning. Even during a single 40 minute regulation signal, SOC estimation uncertainty reaches 1\%. One regulation market profiting mechanism occurs when a BESS owner profits from maintaining available capacity for a potential regulation signal. Given that the profits are proportional to capacity, this could lead to conservative bidding that would cost the owner 1\% of profits. This illustrates the need for BESS control algorithms to introduce a rest period to calibrate the SOC and decrease uncertainty.

We note that at rest, the uncertainty does not decay to zero, but rather levels out to a steady state value, as shown in Fig.~\ref{u_const}. This value results from the dynamic inversion error in the NMC cell. In a different cell chemistry with a flatter OCV-SOC curve, such as Lithium Iron Phosphate (LFP), this model would lead to a much larger error in Eq.~\eqref{ev1}~\citep{pozzato2022core}. The consequence of this error is a large steady state uncertainty at rest, which is consistent with the lack of SOC observability in LFP cells.

For many battery systems, such as residential BESSs or EVs, rest periods are inherent to usage. Large-scale BESSs, however, may be constantly charging or discharging to profit from dynamic electricity prices. A daily built in rest period would allow for sufficient SOC calibration.

\section{Conclusion}
While BESSs can have a significant positive impact on grid stability and demand response, their large capital and maintenance costs mean that maximizing profits from participating in arbitrage or frequency regulation is paramount. Quantifying SOC uncertainty based on known sensor and model errors assists in choosing whether to be conservative or not in planning power trajectories that rely on an accurate initial SOC. In this work, we demonstrate that even concatenating just two 40 minute frequency regulation signals causes a SOC uncertainty of greater than 1\%. A 30 second rest between each signal, however, ensures the uncertainty stays below 1\%. While this is a promising result in simulation, future research will include experimental validation considering temperature variation on a physical BESS to ensure the SOC value reported by the BESS falls within the SOC uncertainty bounds calculated in simulation.

\bibliography{articles}             

\end{document}